\documentclass[sigconf,twocolumn,nonacm]{acmart}

\makeatletter
\def\@ACM@checkaffil{
    \if@ACM@instpresent\else
    \ClassWarningNoLine{\@classname}{No institution present for an affiliation}%
    \fi
    \if@ACM@citypresent\else
    \ClassWarningNoLine{\@classname}{No city present for an affiliation}%
    \fi
    \if@ACM@countrypresent\else
        \ClassWarningNoLine{\@classname}{No country present for an affiliation}%
    \fi
}
\makeatother

\copyrightyear{2023}
\acmYear{2023}
\setcopyright{acmcopyright}
\acmConference[CIDR '23] {Proceedings of the 2023 Annual Conference on Innovative Data Systems Research}{January 8-12, 2023}{ Amsterdam, Netherlands}
\acmBooktitle{Proceedings of the 2023 Annual Conference on Innovative Data Systems Research (CIDR '23), January 8-12, 2023,  Amsterdam, Netherlands}
\acmPrice{15.00}
\acmISBN{978-1-4503-8343-1/21/06}
\acmDOI{10.1145/XXXXXX.XXXXXX}

\usepackage{subcaption}
\usepackage{standalone}
\usepackage{color, colortbl}
\usepackage{listings}
\usepackage{scalerel}
\usepackage{tikz}

\usepackage[vlined,commentsnumbered,ruled]{algorithm2e}
\let\oldnl\nl
\newcommand{\nonl}{\renewcommand{\nl}{\let\nl\oldnl}}

\SetCommentSty{mycommfont}

\def\HiLi{\leavevmode\rlap{\hbox to \hsize{\color{red!20}\leaders\hrule height .8\baselineskip depth .5ex\hfill}}}

\usepackage{amsfonts}
\usepackage{amsmath}
\usepackage{comment}
\usepackage{multirow}
\usepackage{paralist}
\usepackage{paralist}

\usepackage{xcolor}
\definecolor{moonstoneblue}{rgb}{0.45, 0.66, 0.76}
\definecolor{oldlace}{rgb}{0.99, 0.96, 0.9}
\definecolor{mintcream}{rgb}{0.96, 1.0, 0.98}
\definecolor{mintgreen}{rgb}{0.6, 1.0, 0.6}
\definecolor{mistyrose}{rgb}{1.0, 0.89, 0.88}
\definecolor{palegold}{rgb}{0.9, 0.75, 0.54}
\definecolor{palechestnut}{rgb}{0.87, 0.68, 0.69}

\newtheorem{example}{Example}[section]

\newcommand{\anna}[1]{\textcolor{red}{\bf (Anna)~[#1]}{\typeout{#1}}}
\newcommand{\revised}[1]{\textcolor{black}{#1}{\typeout{#1}}}

\newcommand{\sysName}{\textsc{CATER}}
\newcommand{\covidDataset}{\textsc{Covid-19}}
\newcommand{\flightsDataset}{\textsc{Flights}}

\newcommand{\reva}[1]{{\leavevmode\color{black}{#1}}}
\newcommand{\revb}[1]{{\leavevmode\color{black}{#1}}}
\newcommand{\common}[1]{{\leavevmode\color{black}{#1}}}

\begin{document}

\setlength{\abovedisplayskip}{0pt}
\setlength{\belowdisplayskip}{0pt}
\setlength{\abovedisplayshortskip}{0pt}
\setlength{\belowdisplayshortskip}{0pt}

\title{Causal Data Integration}

\author{Brit Youngmann$^1$, Michael Cafarella$^1$, Babak Salimi$^2$, Anna Zeng$^1$}
\email{brity@mit.edu, michjc@csail.mit.edu, bsalimi@ucsd.edu, annazeng@mit.edu}
\affiliation{%
  \institution{$^1$CSAIL MIT, $^2$ University of California San Diego}
}




\begin{abstract}
Causal inference is fundamental to empirical scientific discoveries in natural and social sciences;
however, in the process of conducting causal inference, data management problems can lead to false discoveries. Two such problems are
(i) not having all attributes required for analysis, and
(ii) misidentifying which attributes are to be included in the analysis.
Analysts often only have access to partial data, and they critically rely on (often unavailable or incomplete) domain knowledge to identify attributes to include for analysis, which is often given in the form of a causal DAG.
We argue that data management techniques can surmount both of these challenges. 
In this work, we introduce the Causal Data Integration (CDI) problem, in which unobserved attributes are mined from external sources and a corresponding causal DAG is automatically built. \common{We identify key challenges and research opportunities in designing a CDI system}, and present a system
architecture for solving the CDI problem. Our
preliminary experimental results demonstrate that solving CDI is
achievable and pave the way for future research.
\end{abstract}
\maketitle



\vspace{-2mm}

\section{Introduction}
\label{sec:intro}
Causal inference lies in the heart of empirical research in natural and social sciences and is commonly used in multiple disciplines, including sociology, medicine, and economics \cite{mcnamee2003confounding,henderson2004handbook}. It aims to answer causal queries, such as, ``Does being overweight cause coronary
heart disease -- independent of cholesterol, and diabetes?" or ``Do tobacco advertisements entice adolescents to buy more cigarettes regardless of whether their parents smoke?"
Causal inference enables analysts to answer causal questions about attributes from a dataset, ultimately enabling them to make real-world discoveries. 
Pearl's framework~\cite{pearl2000models}, which we adapt in this work, provides a principled way to causal inference using structural causal models.

Cause-effect questions are designed to determine whether an \emph{exposure} variable (a pain reliever) causes or affects an \emph{outcome} variable (pain). 
To correctly estimate causal effects, one must consider \emph{confounding variables}---variables that influence both the exposure and outcome and thus might distort the association between them.
Otherwise, one might draw perplexing conclusions due to {\em confounding bias}~\cite{pearl2018book} that can lead to overestimation or underestimation of the association between exposure and outcome.
The internal validity of a study heavily relies on how well confounding bias has been addressed. 
However, to determine a sufficient set of confounding variables to account for confounding bias, \emph{background knowledge} about the data generative process is required. This knowledge is often given in the form of a \emph{causal DAG} depicting
causal relationships between the attributes \cite{pearl2000models} (and is wildly used in econometric and social sciences~\cite{henderson2004handbook, mcnamee2003confounding}). Pearl~\cite{pearl2000models} presented sufficient and necessary conditions for identifying\footnote{In causal inference, identifiability typically refers to the conditions that permit measuring causal effect from observed data. Here, we discuss the ability to identify a sufficient adjustment set of variables for accurate causal inference.} the \emph{adjustment set} of variables to include in the analysis, which can be checked against a causal DAG.
However, analysts often lack such a DAG~\cite{glymour2019review}.
While associational assumptions are testable from data,
causal relations cannot, in general, be fully recovered from data~\cite{pearl2018book}.

We identify two fundamental challenges for conducting valid causal inference: First, critical confounding variables may not be included in the data.
Second, even if they are included, the background knowledge required 
to identify the correct adjustment set of variables might be missing. 
We argue that data management techniques and ideas can surmount both of these challenges. 
We illustrate that via the following example:

\emph{
Mary, a data analyst in the WHO organization, aims to estimate the (treatment) effect of a mask policy on the coronavirus mortality rate.
To do this, \reva{she examines a dataset containing Covid-19-related facts in multiple states in the US (an illustration of this dataset is given in Table \ref{tab:example})}, and uses a Causal analysis tool (e.g., \cite{sharma2020dowhy,tingley2014mediation}).
To correctly estimate the effectiveness of face masks, Mary must consider confounding variables.
However, there are confounding variables that are not included in her data. For example, the weather affects people's willingness to wear masks and Covid-19 death rate~\cite{schauer2021analysis}. Moreover, to determine a sufficient set of confounding variables, Mary must use background knowledge of causal relationships she may lack.}

 \begin{table}
	\centering
	
	\scriptsize
		\caption{\reva{Example Input Dataset}.}
			\label{tab:example}
			\vspace{-4mm}
	\begin{tabular} {|p{4mm}|p{8mm}|p{10mm}|p{8mm}|p{10mm}|p{8mm}|}
		\hline
\textbf{\reva{State}} &	
\textbf{\reva{Mask Policy}} & \textbf{\reva{Confirmed Cases}} &\textbf{\reva{New Cases}} & \textbf{\reva{Recovered}} &\textbf{\reva{Death cases}}
	 \\
	
		\hline

 \reva{MA} & \reva{yes}&\reva{121046}&\reva{2740}&\reva{4980}&\reva{109}\\ 
         \reva{FL}  &\reva{yes}&\reva{640978}&\reva{24349}&\reva{25140}&\reva{55}\\ 
    \reva{CA}&\reva{no}&\reva{735235}&\reva{31150}&\reva{42170}&\reva{34}\\

          \reva{SD} &\reva{no}&\reva{15300}&\reva{1791}&\reva{2083}&\reva{49}\\ \hline

	\end{tabular}
	\vspace{-6mm}
\end{table}

In this example, variables missing from the dataset are critical for the analysis to avoid confounding bias. Thus, Mary must \emph{integrate her data} with external data sources. This is a laborious task, typically done using data discovery and integration tools (e.g., \cite{nargesian2018table,zhang2020finding}) by skilled programmers---which is often not the case for social or natural scientists. We argue this is the case for many real-life scenarios.
After augmenting the data to include unobserved variables, the next step is determining a sufficient adjustment set of variables to account for confounding bias.
However, Mary does not have the required background knowledge to do that.

\noindent
\textbf{Our Vision:} 
Our goal is to mine unobserved confounding variables and missing background knowledge needed to answer causal questions. To this end, we introduce the Causal Data Integration (CDI) problem. 
CDI is an effort to augment an input dataset with unobserved variables and background knowledge required for causal analysis.  
To address the challenge of finding unobserved variables, the system mines variables from external sources. 
To overcome the challenge of identifying the right adjustment set of confounders, the system then builds a corresponding causal DAG, ensuring the adjustment set is sufficient for controlling the confounding bias. A CDI system may greatly reduce the manual effort devoted to performing causal analysis, allowing analysts to get all required data and background knowledge quickly and effectively.

\reva{CDI may also be beneficial for the data management community. Causal inference has been shown to be useful for various data management tasks, including query result and classifier explanation \cite{salimi2018bias,alipour2022explaining,youngmann2022explaining}, and hypothetical reasoning \cite{galhotra2022hyper}.  
However, some of these studies make strong assumptions about the underlying causal DAG that may not hold in practice~\cite{mcnamee2003confounding,henderson2004handbook}. For example, \cite{galhotra2022hyper} assumes that the DAG is known, and \cite{salimi2018bias} assumes that all confounding variables are present in the input dataset. To address these issues, a CDI system can be utilized to construct a causal DAG for a given dataset while integrating relevant missing attributes. This not only addresses the limitations of existing studies but also enables the use of causal inference in more complex and realistic scenarios. } 


In this work, we show how a CDI system can be accomplished.

\noindent
\textbf{Measure of Success}: 
\revb{Distinct from the classical data integration task, where success is defined by its ability
to combine data from different sources and provide users with a unified materialized view of them}, the success of a CDI system is defined by the ability to discover all relevant unobserved variables and recover the correct set of confounders for a causal question. 
The CDI task is potentially easier than data integration. However, its impact is greater, as it enables analysts to conduct causal analysis more easily. 
A CDI system enables analysts to include all confounding variables in the analysis; failing to do so may lead to false discoveries, incorrect medical diagnoses, and erroneous conclusions \cite{mcnamee2003confounding,henderson2004handbook}. In our example, a CDI system mines, among others, attributes describing the weather and population density per state, and infers causal relations among the attributes in the augmented dataset. A successful CDI system would assist Mary in identifying the right adjustment set of attributes to account for confounding bias, allowing her to accurately estimate the effect of an enforced mask policy on the mortality rate.

\noindent
\textbf{\common{Challenges \& Opportunities}}:
\common{The first challenge is to augment an input dataset with unobserved variables, ensuring the \emph{output causal DAG is complete} (i.e., there are no unobserved confounding variables). 
Data discovery methods have been extensively researched to identify relevant data sources that can be integrated with an input dataset for attribute extraction~\cite{bogatu2020dataset,castelo2021auctus}. However, these methods are not specifically designed to discover unobserved attributes that are relevant for cause-effect estimation. This presents a unique opportunity to develop dedicated data discovery tools tailored for identifying unobserved confounding variables.} 


\revb{Extracted attributes may contain data quality issues, such as outliers and missing values. These issues may lead to erroneous conclusions, compromising the reliability of causal inference. In particular, missing values pose a significant challenge in causal inference as they may result in \emph{selection bias}~\cite{bareinboim2012controlling}, where certain samples are preferentially excluded from the data. Other data quality issues, such as inconsistent data formats and data duplication, can also impact the accuracy of causal inference analysis. Therefore, in addition to handling missing values, it is crucial to ensure that the system is robust to various data quality issues to ensure the validity of causal inference results.}

\common{The challenge of identifying the correct adjustment set of confounders and constructing a causal DAG is particularly difficult when dealing with hundreds of extracted attributes. A human-driven causal DAG construction (from domain expertise) is infeasible in this scenario, and automatic methods for building a full causal DAG over all attributes may not be feasible either~\cite{glymour2019review}. To this end, we propose the construction of a cluster causal DAG (C-DAG)~\cite{anand2022effect}, which groups related attributes and only specifies causal relationships between clusters of attributes. By doing so, the correct adjustment set of attributes can still be identifiable from this C-DAG. This approach reduces cognitive overhead for analysts and makes causal analysis possible for high-dimensional data, which is currently not feasible for existing tools (e.g., \cite{sharma2020dowhy,tingley2014mediation}). 
However, a main challenge is how to maintain the necessary level of granularity to ensure accurate causal inference. 
The limitations of existing solutions and challenges in constructing the desired C-DAG are discussed, and a limited prototype implementation for solving the CDI problem is presented with preliminary experimental results demonstrating that solving CDI is achievable.}




\section{The CDI Problem}
\label{sec:problem}

\subsection{Data Model and Background}
\label{subsec:data_model}
\revb{The input dataset $\mathcal{D}$ contains the exposure ($T$) and outcome ($O$) attributes. 
A \emph{causal DAG} for $\mathcal{D}$ is a DAG whose nodes are
the attributes and whose edges capture all causal relationships between the attributes \cite{pearl2000models}. Generally, because causes must precede effects, the literature considers causal DAGs rather than graphs \cite{pearl2000models}.
Causal DAGs provide a simple way of graphically representing causal relationships and are particularly helpful in understanding potential sources of bias in causal estimations. It is well-known that background knowledge is required to determine a causal DAG for a given dataset~\cite{pearl2018book}. A causal DAG can only be as good as the background information used to create it~\cite{pearce2016causal}; a DAG is complete and therefore has a causal interpretation only if it contains all common causes of any two variables. In Section \ref{subsec:dag_builder}, we review existing approaches for building a causal DAG for a given dataset and discuss their limitations.}
We assume the analyst is interested in investigating a causal relationship between $T$ and $O$. She may wish to estimate, e.g., \emph{the direct, indirect, or total effect}~\cite{pearl2000models}.
The total effect of $T$ on $O$ is the total extent to which $O$ is changed by the $T$. It 
is equal to the sum of the direct and indirect effects. Causal questions about total and (in)direct effect could be answered by identifying the right set of \emph{confounding} (that influence both $T$ and $O$) and \emph{mediating} (that mediate the relationship between $T$ and $O$) variables. Such variables could be identified using graphical criteria (e.g., the backdoor criterion \cite{pearl2000models}) that can be checked against a causal DAG.


\begin{figure}
\centering

\includegraphics[scale = 0.24]{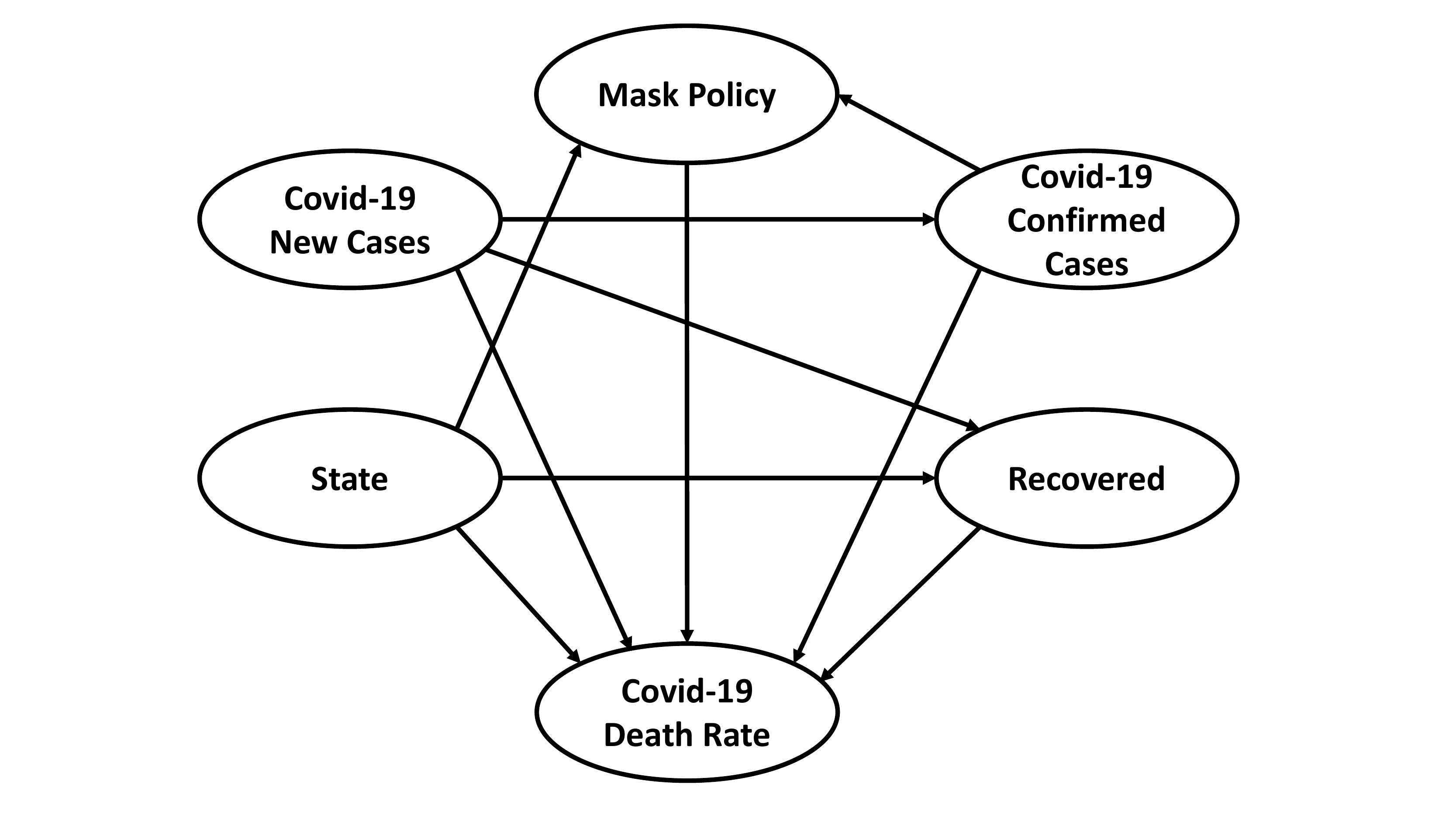}
\vspace{-6mm}
\caption{\reva{Example causal DAG for Table \ref{tab:example}}} \label{fig:causal_dag}
\vspace{-6mm}
\end{figure}

\begin{example}
\label{ex:causal_dag}
\reva{In our example (Table \ref{tab:example}) $T$ is } \verb|mask policy|, \reva{and $O$ is } \verb|death cases|. \reva{A corresponding causal DAG is shown in Figure \ref{fig:causal_dag}. To obtain a reliable estimate of the total effect of $T$ on $O$, it is crucial to adjust for all relevant confounding variables. However, mistakenly assuming that the DAG contains all relevant information, the adjustment set only includes the attributes} \verb|state| and \verb|confirmed cases|. \reva{As is common in causal inference, missing confounding variables can lead to biased estimates of causal effects. Thus, it is necessary to consider external data sources to incorporate missing variables. 
}
\end{example}

\vspace{-2mm}
\subsection{Problem Formulation \& Challenges}
\label{subsec:challenges}
Given an input dataset, the system mines unobserved confounding attributes relevant for a given causal estimation, yielding an augmented dataset $\widehat{\mathcal{D}}$. It then maps the attributes in $\widehat{\mathcal{D}}$ into a causal DAG, to enable analysts to determine the adjustment set. We next outline key challenges in designing such a
system:\\
\textbf{(1) Completeness:} 
\common{Unobserved variables can dramatically affect the quality of causal analysis~\cite{lindmark2021sensitivity,pearl2018book}.
Thus, a first challenge is to \emph{ensure the generated DAG is complete}, i.e., there are no unobserved confounding variables w.r.t. the provided data sources. Our goal is to ensure that all relevant information can be automatically extracted from the provided data sources. However, completeness, in general, cannot be guaranteed since relevant variables may exist outside of the provided sources of data.}\\
\textbf{(2) Robustness:}
\revb{The robustness of a CDI system is determined by its ability to handle various data quality issues that may arise during the data integration process. As in general data integration, there are multiple data quality issues that a CDI system has to face, including missing values and outliers. 
The impact of such issues on causal inference is particularly significant. Unlike in traditional machine learning, where data quality issues may only result in a slight drop in accuracy, these issues can lead to false discoveries in causal inference if not handled properly. This is because if these issues are not randomly distributed across the exposure groups, they can significantly impact the validity of the resulting conclusions. One common issue that can significantly impact causal inference is missing data. Extracted attributes may contain missing values, which can lead to selection bias and erroneous conclusions. Thus, a key challenge is to ensure the system is robust to---and effectively handles ---various data quality issues.}\\
\textbf{(3) Conciseness:} \revb{We can potentially extract hundreds of attributes from given data sources, which can lead to a complex, high-dimensional causal DAG. To overcome this, the CDI system should aim to reduce users' cognitive overhead in interpreting the DAG. It is also crucial to ensure the scalability of causal analysis, as existing tools are not feasible for high-dimensional data~\cite{sharma2020dowhy,tingley2014mediation}.
One approach is to ensure that related attributes are grouped into clusters while still ensuring that the correct set of confounders is identifiable. Thus, a third challenge is to ensure the output DAG is concise. The CDI system should be able to construct a causal DAG that effectively captures the causal relationships between attributes while maintaining the necessary level of granularity to ensure accurate causal inference. This will enable analysts to efficiently handle high-dimensional data while facilitating information elicitation about the validity of the assumptions used for drawing causal inferences.}


\section{The CDI System}
\label{sec:system}

	

\revised{An analyst provides to the system a dataset $\mathcal{D}$ containing the exposure and outcome and external data sources. 
The system mines unobserved variables from the data sources, yielding an augmented dataset $\mathcal{D'}$. It then maps the variables in $\mathcal{D'}$ into a causal DAG $\mathcal{G}$. The analyst can then download the augmented dataset, and identify the adjustment set of variables using $\mathcal{G}$.
Last, she uses a causal analysis tool to perform her analysis.
Such tools take as input a dataset (contains all required attributes) and assume the analyst knows which variables to include in the analysis.} \common{Our system operates as a 3-steps pipeline. First, the \emph{Knowledge Extractor}, extracts candidate attributes to handle the \emph{completeness} challenge. The \emph{Data Organizer} then handles data quality issues to account for the \emph{robustness} challenge. Last, the \emph{Causal DAG Builder} builds a clustered casual DAG to handle the \emph{conciseness} challenge. We next present the key challenges and research opportunities for each system component. }


\begin{table}
\caption{\reva{Extracted attributes from different sources}.}
 \label{tab:extracted_att}
 \vspace{-5mm}
	{\scriptsize
		\begin{center}
			\begin{tabular}[b]{|l|l|l|}
				\multicolumn{3}{c}{\reva{Extracted Attributes from US Open Data}}\\ \hline
				\reva{State} & \reva{Population size} & \reva{Population density}      \\ \hline
    \reva{MA}&\reva{6,981,974}&\reva{901}\\
    \reva{FL}&\reva{22,244,823}&\reva{402}\\
    \reva{CA}&\reva{39,029,342}&\reva{254}\\
    \reva{SD}&\reva{909,824}&\reva{12}\\

				\hline
			\end{tabular}
			\quad
			\begin{tabular}[b]{|l|l|l|l|l|}
				\multicolumn{5}{c}{\reva{Extracted Attributes from DBPedia}}\\
				\hline
				\reva{State}   & \reva{Governor} & \reva{snow inch} & \reva{Avg temp.} & \reva{Min temp.}   \\ \hline
      \reva{MA}&\reva{Maura Healey}&\reva{51.05}&\reva{48.14}&\reva{23}\\
    \reva{FL}&\reva{Ron DeSantis}&-&\reva{71.8}&\reva{70}\\
    \reva{CA}&\reva{Gavin Newsom}&-&\reva{61.17}&\reva{54}\\
    \reva{SD}&\reva{Kristi Noem}&\reva{37.43}&\reva{45.54}&\reva{34}\\
			
				\hline
				
			\end{tabular}
		\end{center}
	}
	\vspace{-5mm}
	
\end{table}

\subsection{The Knowledge Extractor}
\label{subsec:knowledgeextraction}
This module receives a table $\mathcal{D}$ (containing $T$ and $O$) and sources of knowledge.
From each source, it extracts attributes representing additional properties of entities from $\mathcal{D}$. A successful Knowledge Extractor mines all unobserved variables that should be included in the analysis. 
\revb{The primary goal of the \emph{Knowledge Extractor} is to extract all relevant variables from the provided data sources to enable reliable causal inference. In causal inference, the \emph{unconfoundedness assumption}~\cite{pearl2009causal} (the assumption that all variables affecting both the exposure $T$ and the outcome $Y$ are observed and can be controlled for) is a key consideration. This assumption is similar to the closed-world assumption in database in that it assumes that the database is complete. However, since complete knowledge of all variables affecting both $T$ and $O$ is often unattainable, the focus should be on extracting all relevant variables from the provided data sources. The module is designed to ensure that as much relevant information as possible can be automatically extracted from the provided sources, and its capabilities can be expanded over time to incorporate new sources or refine the extraction and integration process.} We next provide an example of how it works with specific data sources.



\noindent
\textbf{Data Lakes}:
There has been unprecedented growth in the volume of publicly available data (e.g., the US Federal Open Data~\cite{usopen} and the Federal Reserve Economic Data~\cite{fred}). We can rely on the rich body of work on data discovery~\cite{bogatu2020dataset,castelo2021auctus} to find
relevant data sources that can be aligned with
the input table for attribute extraction. However, existing methods use notions of relevance for discovery \cite{zhang2020finding, nargesian2018table,zhu2019josie}, whereas we search for unobserved confounding variables relevant to cause-effect estimations.
Recent work has proposed solutions that can discover joinable datasets with column(s) correlated with a given target column and an input table~\cite{esmailoghli2021cocoa,santos2021correlation}. However, future research is required to extend these techniques for extracting candidate confounding attributes from tabular data, which can further improve the completeness of the extracted information.


\noindent
\textbf{Knowledge Graphs}: 
There are multiple general-purpose (e.g., DBpedia~\cite{dbpedia})  
or domain-specific (e.g., for medical proteomics \cite{santos2022knowledge}, or protein discovery \cite{mohamed2020discovering}) KGs that act as central storage for data. 
To extract attributes from a KG, 
we may map values that appear in $\mathcal{D}$ to their corresponding entities in the KG $\mathcal{G}$, using named entity disambiguation tools (e.g., \cite{parravicini2019fast}). 
Next, given an entity, one can extract all of its properties, building the universal relation~\cite{fagin1982simplied} 
out of all derived relations. 
One of the strengths of a KG is that most of the attributes are already reconciled.
To extract more attributes, one may "follow" links in $\mathcal{G}$ (i.e., extract the properties of values which are entities in $\mathcal{G}$ as well). 
Future work will identify which links in a KG are worth following to extract relevant confounders.

\noindent
\textbf{Unstructured Data}: \common{Unstructured data such as text and images can be a rich source for identifying missing confounding variables. 
For example, medical records often contain unstructured text such as physician notes, which can provide insights into patients' health status and potential confounding variables. Similarly, medical images can contain important features that may impact treatment outcomes. However, unstructured data pose a challenge to causal analysis due to their lack of organization. Feature extraction techniques can be applied to extract meaningful information~\cite{zeng2022uncovering}. For example, sentiment analysis can be applied to text to extract relevant features, and computer vision methods can be used to extract image parts. This can 
 improve the accuracy of causal analysis.
 Future work could develop methods for extracting interpretable features from unstructured data tailored for causal inference.}

\begin{example}
\label{ex:knowledge_extractor}
\reva{The \emph{Knowledge Extractor} extracts potential confounding variables from multiple knowledge sources. Specifically, it extracts the population
density and size by state from the US Open data lake by joining the input table with a table containing statistics on population (based on state names). From DBpedia, it extracts available properties of each state (by aligning the state name with its entity in DBpedia), such min temperature, amount of snow, and governor. A visualization of extracted attributes is given in Table \ref{tab:extracted_att}.} 
\end{example}

\textit{\textbf{\common{Challenges \& Opportunities}}}:
\common{The \emph{Knowledge Extractor} faces the challenge of identifying relevant unobserved attributes to be added for any arbitrary input dataset. The module must operate correctly even in cases of value mismatches, such as when a state name is given explicitly or as a shortcut. To tackle this challenge, researchers can rely on a rich body of work on entity linking and disambiguation \cite{demartini2013large,lin2020kbpearl,dredze2010entity}. Another challenge is to ensure that only relevant attributes are extracted and to avoid the curse of dimensionality. To achieve this, we plan to draw inspiration from first principles, such as an information-theoretic approach or approaches based on the maximum likelihood principle.} 

\vspace{-2mm}
\subsection{The Data Organizer}
\label{subsec:data_organizer}
\revb{The \emph{Data Organizer} receives the extracted attributes and outputs an augmented dataset. Extracted attributes may
contain data quality issues, such as missing values and duplicated data, which can compromise the reliability of causal analysis. A successful implementation of
this module would handle all data quality issues and ensure the system is robust to data quality issues.} 
\common{We next discuss several example failure modes the \emph{Data Organizer} have to face}: \\
\textbf{Functional Dependencies}: 
\common{We search for attributes affecting both the exposure $T$ and the outcome $O$. The presence of logical dependencies can hinder this process, as they can obscure the true relationships between them. In particular, causal inference literature assumes that the underlying distribution is \emph{strictly positive}~\cite{pearl2000models}. 
This assumption breaks down in the presence of logical dependencies: Assume we have an attribute $E$ with the functional dependency $E {\Rightarrow} T$. This means that when conditioning on $E$, there is no association between $T$ and $O$, 
regardless of the choice of $O$. 
Namely, the existence of $E$ violates the strict positivity assumption.  
A straightforward solution is to discard such attributes~\cite{salimi2018bias}. Another solution is to group attributes with functional dependencies, treating them as a single node. We plan to develop a principled approach to discovering causal relations with the presence of functional dependencies. } 
\textbf{Missing Values}:
\common{
Inferring causal relations requires recovering the probability distributions of each extracted attribute~\cite{glymour2019review}. 
But since extracted attributes may contain missing values, we must ensure that those probabilities are \emph{recoverable} to overcome potential selection bias. 
Inverse Probability Weighting (IPW) is
a commonly used technique to overcome selection bias~\cite{seaman2013review}. In IPW, we restrict the attention only to complete tuples, but more weight is given to some tuples. 
A main challenge is to define sufficient conditions to detect cases where probabilities are not recoverable (and thus selection bias may occur). In such cases, there is a need for a principled approach to assigning weights to tuples whose values were extracted (following the IPW approach) to be used to discover causal relations correctly. This can be done by extending the ideas proposed in \cite{youngmann2022explaining}. }
\noindent
\textbf{Complex Table Relations}: 
\common{Another challenge of the \emph{Data Organizer} is to handle complex table relations.} Extracted attributes may have one/many-to-many relations with values in the input table. 
We aim to generate a single, augmented table. 
A straightforward solution is to aggregate multi-valued attributes; another possible direction is to develop representation learning techniques tailored for causal inference~\cite{bengio2013representation}.  
To accommodate many-to-many relations, one may use similar ideas as proposed in~\cite{salimi2020causal}.

\textit{\textbf{Challenges \& Opportunities}}:
\common{Other data quality issues, such as inconsistencies, and outliers, can also compromise the reliability of causal inference and lead to erroneous conclusions if not handled properly. This is because if these issues are not randomly distributed across the exposure groups, they can lead to bias that may significantly impact the validity of the resulting conclusions. Merely using existing data-cleaning techniques (e.g., \cite{chu2016data,ilyas2019data}) may not be sufficient in the context of causal inference, as they are not designed to validate that bias is resolved. Thus, there is a need for innovative techniques that can effectively address these data quality issues to ensure accurate and reliable causal analysis. This is particularly important in high-stakes domains, such as healthcare, where erroneous conclusions can have severe consequences.}


\begin{example}
\label{ex:data_org}
    \reva{The \emph{Data Organizer} 
    identifies the functional dependency between} \verb|state| \reva{and} \verb|governor| \reva{and drops the} \verb|governor| \reva{attribute. For attributes containing missing values (e.g.,} \verb|snow inch|), \reva{it checks if selection bias might occur. If so, it computes the weights for the existing values required for the analysis. It then joins the input dataset with the extracted attributes, yielding an augmented dataset.}
\end{example}



\begin{figure}
\centering

\includegraphics[scale = 0.24]{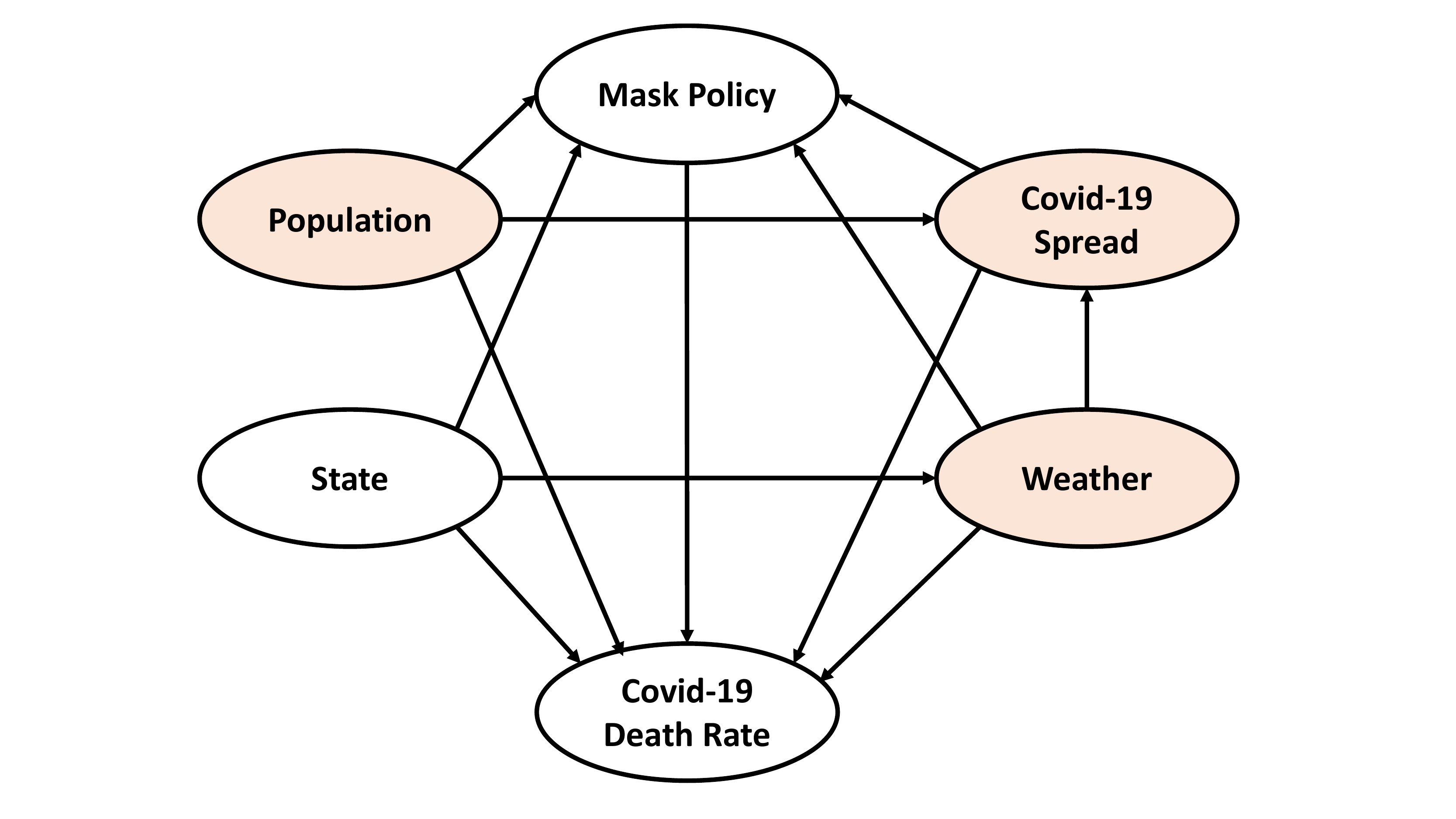}
\vspace{-6mm}
\caption{\reva{The generated C-DAG. Nodes representing a cluster of nodes are colored pink.}} \label{fig:c_dag}
\vspace{-5mm}
\end{figure}

\vspace{-2mm}
\subsection{The Causal DAG Builder}
\label{subsec:dag_builder}

This module receives the augmented dataset and outputs a \emph{concise causal DAG}.
Any successful \emph{Causal DAG Builder} should output a comprehensible and concise causal DAG while ensuring that the correct adjustment set of attributes is identifiable from that DAG.

\vspace{1mm}
\noindent
\revb{\textbf{Desiderata}: The desiderata for a desired causal DAG is summarized as follows. First, there is a need to limit the number of nodes in the output DAG. This is critical to ensure that causal analysis is possible for high-dimensional data (which is not feasible for existing tools \cite{sharma2020dowhy,tingley2014mediation}) and will reduce analysts' cognitive overhead 
in interpreting the DAG. Second, to ensure the output DAG is comprehensible, only semantically-related attributes (e.g.,} \verb|avg temp|, \verb|snow inch|) \revb{should be grouped and represented as a single node in the DAG. Last, we must ensure that the output DAG enables identification of the right adjustment set of confounders for accurate causal inference. }

\vspace{1mm}
\noindent
\textbf{Clustered Causal DAG}: \revb{To address the above desiderata, we propose to build a \emph{cluster causal DAG} (C-DAG)~\cite{anand2022effect}.  
A C-DAG provides only a partial specification of causal relationships among attributes, alleviating the requirement of specifying all causal relationships. In a C-DAG, some attributes are grouped, and only the causal relationships between clusters of attributes are specified. 
Previous work~\cite{anand2022effect}
showed that cause-effect computations can be done directly from a C-DAG. In the output C-DAG, only semantically-related attributes should be clustered, and the number of nodes should be limited. A remaining open question is how to ensure that the correct adjustment set of confounding attributes is identifiable since, by grouping attributes, we lose information about their inter-relations. We next discuss the main challenges in building a desired C-DAG. }

\noindent
\textbf{\textit{Identifying Causal Relationships}}: 
Causal discovery is a well-studied problem~\cite{glymour2019review,spirtes2000causation,shimizu2006linear,wiering2002evolving,zhu2019causal}, whose goal is to infer causal relations among a set of attributes. Existing solutions can be split into two main approaches.
Data-centric methods infer a causal DAG based on data properties~\cite{glymour2019review}.
Though it is well-known that background knowledge is required to determine causal relations~\cite{pearl2018book}, 
a causal DAG can be inferred from the data
under some assumptions 
\cite{glymour2019review,chickering2002optimal} (e.g.,
sufficiency,
faithfulness). 
However, such assumptions
may not hold in practice and therefore limit the applicability of these algorithms.
Further, such algorithms may not capture causal relations that are not present in the data
and ignore available semantic information.
Text-mining approaches (e.g., \cite{hashimoto2019weakly,heindorf2020causenet,hassanzadeh2019answering}), on the other hand, infer causal relations among concepts by extracting claims of causal relationships from text documents. 
Other promising approaches are large language models such as ChatGPT and GPT-3~\cite{brown2020language} that have been trained to discover causal relations.  
The strength of such methods is in identifying cause-effect relations between seemingly-independent attributes or relations that otherwise could not be found in the data.
However, they lack any concrete data on which to validate causal relations and may fail to distinguish between direct and indirect effects, resulting in graphs containing redundant edges. 
Further, such methods are also sensitive to the quality of attribute names and cannot be fully trusted.

\common{We argue that a hybrid approach that merges the two approaches is required to overcome their shortcomings. Such an approach uses both data-centric and text-mining solutions to infer causal relations. 
However, achieving high
accuracy in determining causal relationships is still a
challenge for CDI; it requires substantial in-domain knowledge. A possible improvement is to use human-in-the-loop tasks, minimizing human effort while maximizing accuracy.}


\noindent
\textbf{\textit{Grouping Attributes}}: 
\common{We aim to group semantically related attributes and lift low-level information (e.g.,} \verb|temp|, \verb|snow|) \common{into higher-level concepts} (\verb|weather|). \common{Semantic similarity between attributes can be measured using embedding techniques (e.g.,~\cite{mikolov2013efficient}), or ontological relationships (e.g.,~\cite{kashyap1996semantic}). 
To allow analysts to reason about the C-DAG, we must further assign meaningful topics to the obtained clusters.  %
This can be done using recent advances in
topic modeling or zero-shot topic classification methods (e.g.,~\cite{vayansky2020review}).}

\noindent
\textbf{\textit{Identifiability}}: 
\common{Besides causal relationships, a causal DAG specifies conditional independencies among the attributes~\cite{pearl2000models}. To ensure the right adjustment set of confounding attributes is identifiable from the C-DAG, the set of conditional independencies among attributes induced by the full causal DAG (where attributes are not clustered) should hold in the C-DAG (and vice versa). But since we group attributes, some of the conditional independencies may not be recoverable from the C-DAG. While there is a rich body of work on graph summarization~\cite{zhang2010discovery,tian2008efficient,kumar2018utility,ko2020incremental}, a tailored causal solution is required to ensure that all relevant conditional independencies are recoverable from the output C-DAG. Another open question is whether a single C-DAG is sufficient to identify the adjustment sets for multiple cause-effect estimations. In Section \ref{sec:poc} 
we propose a best-effort algorithm to build a C-DAG.}

\begin{table*}
	\centering
	\captionsetup{justification=centering}	
	\scriptsize
		\caption{Quality Evaluation.}
			\label{tab:quality}
			\vspace{-12px}
	\begin{tabular} {|p{27mm}|p{12mm}|p{12mm}|p{9mm}|p{7mm}|p{7mm}|p{9mm}|p{7mm}|p{7mm}|p{15mm}|}
		\hline
\textbf{Dataset}&	\textbf{Baseline}&\textbf{|E|} &
\multicolumn{3}{c|}{\textbf{Inclusion of Directed Edges}}&\multicolumn{3}{c|}{\textbf{Absence of Edges}}&\textbf{Direct Effect}\\
\cline{4-9}

&&&

\textbf{Precision}& \textbf{Recall}&\textbf{F1}&\textbf{Precision}&\textbf{Recall}&\textbf{F1}&
	 \\
		\hline
\multirow{6}{*}{\flightsDataset\ ($|V| = 9, |E| = 17$)}&
\sysName&25&0.94&0.64&\textbf{0.74}&0.62&0.72&\textbf{0.66}&\textbf{0.04}\\
&	GES&33&0.94&0.48&0.64&0.60&0.43&0.50&0.2\\
&	LinGAM&23&0.70&0.52&0.60&0.65&0.41&0.50&0.2\\
&	PC&27&0.70&0.44&0.54&0.62&0.42&0.49&0.2\\
&	GPT-3 Only&63&1.0&0.27&0.43&0.16&1.0&0.28&0.07\\
&	FCI&39&0.70&0.30&0.43&0.40&0.30&0.35&0.15\\

		\hline
		\hline
		\multirow{6}{*}{\covidDataset\ ($|V|=11, |E|=23$)}&	\sysName&27&0.74&0.63&\textbf{0.68}&0.59&0.61&\textbf{0.6}&\textbf{0.01}\\
&	GPT-3 Only&82&0.91&0.26&0.4&0.25&0.78&0.38&0.02\\
&	GES&16&0.26&0.37&0.31&0.49&0.31&0.38&0.05\\
&	PC&13&0.17&0.31&0.22&0.49&0.3&0.37&0.05\\
&	FCI&13&0.13&0.23&0.17&0.26&0.24&0.25&0.05\\
&	LiNGAM&1&0&0&0&0.63&0.42&0.5&0.05\\

		\hline

	\end{tabular}
	\vspace{-4mm}
\end{table*}

\begin{example}
\label{ex:causal_dag_builder}
\reva{In our example (see Figure \ref{fig:c_dag}), the \emph{C-DAG builder} groups low-level related attributes (e.g.,} \verb|pop density|, \verb|pop size|) \reva{into clusters associated with relevant topics (e.g.,} \verb|population|), \reva{and infers causal relations among the attribute clusters.} 
\end{example}


\section{PROOF OF CONCEPT IMPLEMENTATION}
\label{sec:poc}
We have built \sysName, 
a limited, proof-of-concept prototype of our CDI system. Our code and datasets are available at \cite{code}.
We next review some early experimental results. 
Though our experiments are limited, our results are highly promising, showing that solving CDI is achievable. 



\noindent
\textbf{Implementation Details}:
\revb{In \sysName, the \textit{Knowledge Extractor} extracts attributes from DBpedia~\cite{dbpedia}. The \textit{Data Organizer} discards attributes that have functional dependencies with the exposure or outcome, following~\cite{salimi2018bias}. 
We detect and handle selection bias in extracted attributes using similar ideas as proposed in~\cite{youngmann2022explaining}. 
The \textit{C-DAG Builder} first groups attributes using VARCLUS~\cite{sarle1990sas} and assigns topics to clusters using GPT-3~\cite{brown2020language}. It then populates the graph
edges according to templated query responses from GPT-3, and prunes redundant edges according to the PC algorithm~\cite{spirtes2000causation}.}
This algorithm may result in a graph containing cycles; however, we report that \sysName\ yields DAGs in examined cases. 

\noindent
\textbf{Datasets}: We examine two datasets: 
\covidDataset~\cite{covid}, which includes information such as
number of confirmed/death/recovered Covid-19 cases worldwide; and
\flightsDataset~\cite{flights}, which contains transportation statistics of domestic flights in the USA.
Our pipeline was executed end-to-end in 645 and 304 seconds
for \flightsDataset\ and \covidDataset, resp.

\noindent
\textbf{Examined Scenarios}:
We use \sysName\ to estimate the direct effect of an exposure variable on an outcome (when not mediated through mediator variables)---a common casual analysis task \cite{pearl2000models}. 
In \covidDataset, we estimate the direct effect of a \verb|country| on \verb|Covid-19 death rate|, and in \flightsDataset, we estimate the effect of \verb|departure city| on \verb|flight delays|. 
The goal is to identify mediator variables that should be included in the analysis, as many such variables are not included in the datasets (e.g., \verb|weather|, \verb|pop. density|).
\emph{In both cases, the direct effect should be equal to zero (there is no direct effect). Failing to identify mediators correctly will result in incorrectly estimating the direct effect and falsely concluding that there is one.}

\noindent
\textbf{Baseline Methods}: 
We picked our current best configurations for the \textit{Knowledge Extractor} and the \textit{Data Organizer}.
\revised{The node clusters and their assigned topics are the same across all baselines.} We 
evaluate different approaches to infer causal relations, demonstrating the superiority of a hybrid approach over existing solutions.
Our baselines include data-centric algorithms: 
\textbf{PC}~\cite{spirtes2000causation}, \textbf{FCI}~\cite{spirtes2000causation}, \textbf{GES}~\cite{chickering2002optimal} and \textbf{LiNGAM}~\cite{shimizu2006linear}, 
and a text-mining approach (asking templated queries to \textbf{GPT-3}~\cite{brown2020language}).
To serve as \textbf{ground truth}, we gather domain expert knowledge C-DAGs. For instance, 
airline carriers and weather are causes for flight delays~\cite{stat}. 

\noindent
\textbf{Quality Metrics}: 
We measure the quality of generated C-DAGs w.r.t. ground truth, in two ways: (1) directed edge presence or absence, and (2) established direct effect between $T$ and $O$. 
The precision, recall, and F1 scores for presence/absence edges are reported; 
High F1 scores indicate alignment with the ground truth.
We report the direct effect between $T$ and $O$ described by each C-DAG. 
\revised{A score close to 0 indicates that all mediator attributes are identifiable from the C-DAG}.

\noindent
\textbf{Results Summary}:
As illustrated in Table \ref{tab:quality}, \sysName\ achieved for both datasets 
the highest F1 scores and the lowest direct effect values.  
Further, in both scenarios, \sysName\ identified the same sets of mediator variables as those identified in the ground truth.
This promising result illustrates that even a naive hybrid approach can improve upon existing causal discovery solutions.

Performance of GPT-3-Only closely trails \sysName's performance, especially in \covidDataset.
Along with the ground truth and \sysName, GPT-3-Only captured many causal relations which data-centric methods were not able to,
such as \verb|climate| ${\rightarrow}$ \verb|spread of Covid-19|. 
However, it outputs graphs with notably more edges than the ground truth.
Consequently, these graphs
are far from being DAGs
(in \covidDataset, there is 2-cycle between \verb|economy| and \verb|population size|).
It is also unable to distinguish between direct and indirect effect.
For instance, GPT-3 and ground truth agree that
\verb|population size| $\rightarrow$ \verb|spread of| \verb| Covid-19| $\rightarrow$ \verb|Covid-19| \verb|death rate|, but GPT-3 also includes the edge \verb|population|  \verb|size| ${\rightarrow}$ \verb|Covid-19 death rate|).

Of the remaining
baselines, GES yielded the highest F1 scores, since it is tolerant to relaxation of the sufficiency assumption. 
Data-centric baselines were able to capture causal relationships that \sysName. For example, in \flightsDataset, these baselines recognized that \verb|origin airport| does not directly cause \verb|departure delay|; attributes like \verb|temp| and \verb|carrier| mediate the relationship.
However, these methods fell behind in discerning mediator variables.
All failed to identify any mediator in \covidDataset\. 
Even with high F1 scores on \flightsDataset, all failed to find any mediator. 
This implies that not all causal relations can be discerned from data alone.

\vspace{-2mm}
\section{Limitations}
\label{sec:conc}


\reva{We note the following limitations of our proposed CDI system: 
First, the quality of the generated C-DAG may
be affected by factors such as the quality of extracted data
(e.g., incorrect values) and black-box components (e.g., the entity linker used to link values from the input dataset to their corresponding entities in a knowledge graph).
Second, the generated C-DAG may
not be complete, meaning that not all real-life confounding
attributes were observed and extracted, and thus the unconfoundedness assumption is violated. In particular, datasets containing information about people often lack properties about their age or ethnicity, which are common confounding variables. 
Finally, the goal of a CDI is to assist analysts in identifying a correct adjustment set to be controlled for to answer a causal question. While methods for identifying confounding variables through causal DAGs using, for example, backdoor and front-door criteria are well understood, they all rely on the availability of full causal DAGs~\cite{pearl2009causality}. However, in practice, full causal DAGs are often not available~\cite{glymour2019review}. Specifically, in our framework, in which we dynamically integrate data with external sources,
it is not always reasonable to assume that a full causal DAG can be automatically discovered. This is because determining causal relationships oftentimes requires substantial in-domain knowledge.}

\clearpage


\small
\bibliographystyle{ACM-Reference-Format}
\bibliography{vldb_sample}

\end{document}